\documentclass{PoS}
\usepackage{wrapfig,lineno}
\title{Public engagement as a scientific tool to implement multi-messenger 
strategies with the Cosmic-Ray Extremely Distributed Observatory}

\ShortTitle{Public engagement in CREDO}

\author{\speaker{Piotr Homola} for the CREDO Collaboration\thanks{The 
CREDO system was developed with support by (alphabetically)
Academic Computig Center Cyfronet AGH-UST, ASTERICS, 
Cracow University of Technology, International Visegrad Fund, and Institute of Nuclear Physics 
Polish Academy of Sciences. Full author list: http://credo.science/publications}\\
Institute of Nuclear Physics Polish Academy of Sciences, Krak\'ow, Poland\\
E-mail: \email{Piotr.Homola@ifj.edu.pl}}

\abstract{The Cosmic-Ray Extremely Distributed Observatory (CREDO) uses the hunt for particle cascades 
from deep space as a vehicle for a unique ``bottom-up'' approach to scientific research. 
By engaging the non-specialist public of all ages as ``citizen scientists'' we create opportunities 
for lifelong learning for individuals as well as for cooperation and the sharing of common educational 
tools amongst institutions. The discoveries of these citizen scientists will feed directly into a pioneering 
new area of scientific research oriented on Cosmic Ray Ensembles (CRE). The detection (or non-detection) of 
such particle groups promises to open up a new method for exploring our universe, and a new channel on the 
multi-messenger stage, oriented on both astro- and geo-investigations. The opportunities this would create 
for cross-disciplinary research are significant and beneficial for individuals, networks of institutions 
and the global communities of both professional scientists and science enthusiasts.}

\FullConference{The New Era of Multi-Messenger Astrophysics - Asterics2019\\
		25 - 29 March, 2019\\
		Groningen, The Netherlands}

\begin{document}

\section{Introduction}
\label{sec:intro}

\begin{wrapfigure}{r}{0.5\textwidth}
  \begin{center}
    \vspace{-0.4cm}
    \includegraphics[width=0.48\textwidth]{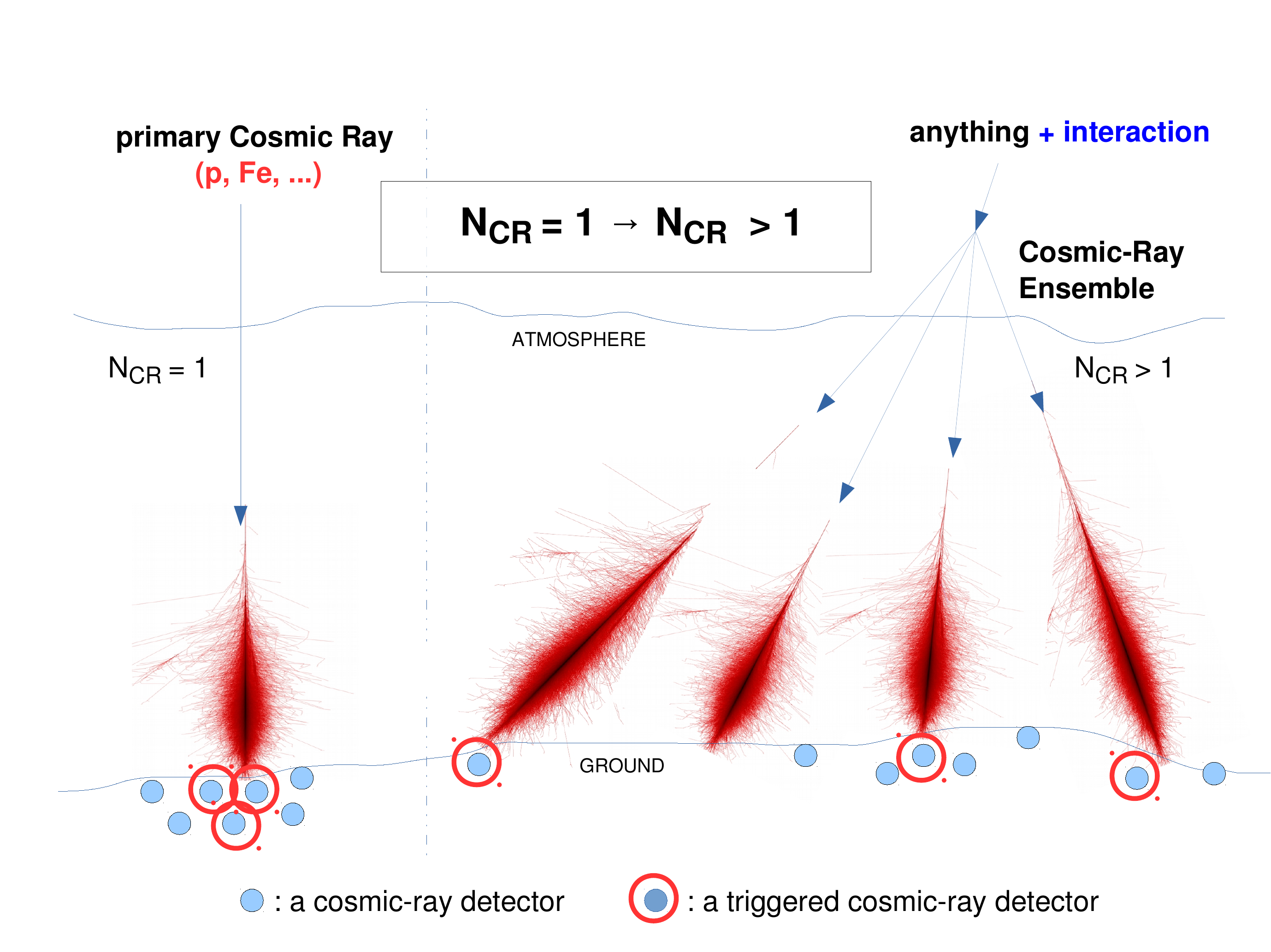}
    \end{center}
  \caption{A generalization of cosmic-ray research 
by admitting Cosmic-Ray Ensembles as a subject of observation.}
\label{fig:ngt1}
\end{wrapfigure}
The Cosmic-Ray Extremely Distributed Observatory (CREDO) aims at searching for the yet
not checked multi cosmic-ray signatures that are composed of many air showers and individual
particles arriving simultaneously to the Earth, so-called cosmic-ray ensembles (CRE), as illustrated
in Fig.~\ref{fig:ngt1}. The signals from CRE might be spread over very large surfaces which might make them hardly
detectable by the existing detector systems operating individually. 
On the other hand, if these detector systems operate under a planetary network, as proposed by CREDO, the chances for
detection of CRE, even as zero-background events, will naturally increase. 
The components of individual CRE might have energies that
span practically the whole energy spectrum of cosmic rays. Thus, all the cosmic-ray detectors working
in this range, beginning from smartphones and pocket
scintillators, through numerous larger educational detectors and
arrays to the professional infrastructure capable of registering cosmic rays as a signal or as a 
background, can contribute to a common effort towards a hunt for CRE.
Since a planetary scale and massive, geographically spread data acquisition with even small
detectors such as smartphones are essential for the CREDO mission, the CREDO Collaboration considers an active
engagement in the project of non-expert science enthusiasts as a methodological must: beneficial for
both for the project and for the contributors, and making science experts and non-professionals just one community, working
hand-in-hand towards common goals. The
unprecedented cosmic-ray data set to be collected and processed by CREDO will require a continuous 
overlook and vigilance of humans, many of us. Thus public engagement must be planned on many levels: from passive data
taking with smartphones, through simple data monitoring, mining and analysis via Internet, to more
advanced activities. In practice, it means that CREDO will offer long-lasting educational and
developmental paths for individuals who became amazed after their first, very close and active
contact with science, and who will be ready for more. If the scale of the CREDO network is as large
as planned, our format must bring benefits not only to the whole science community but also to the
society. Universally useful educational and mind formation opportunities together with the
availability of indirect and immediate participation in a top-science project for the so-far-excluded
countries, regions and individuals, should contribute to the overall, sustainable civilizational
development of society.

\section{Public engagement in CREDO: for science, through contributing to science}
\label{sec:engagement}

\begin{wrapfigure}{r}{0.5\textwidth}
  \begin{center}
    \vspace{-0.5cm}
    \includegraphics[width=0.48\textwidth]{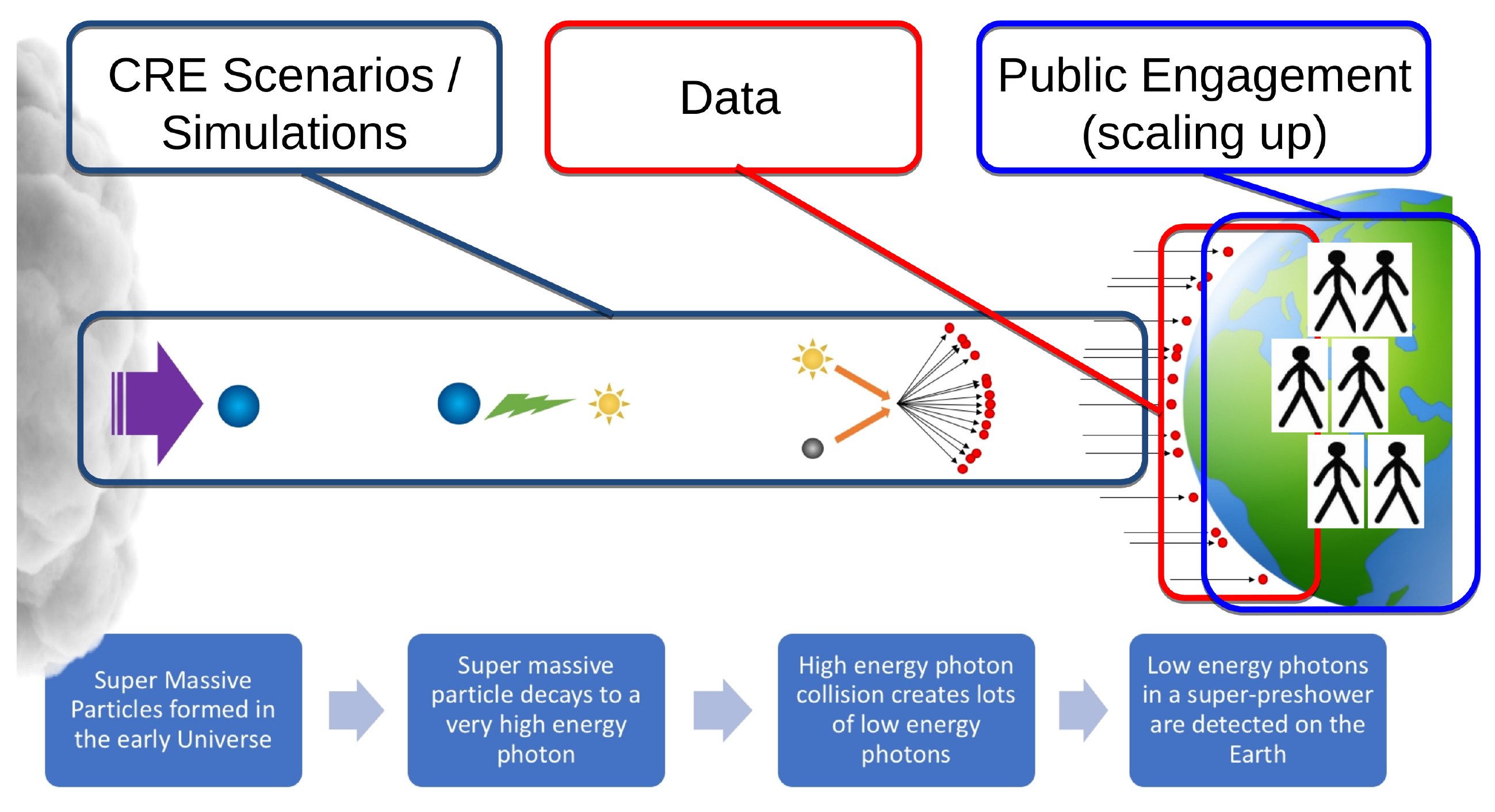}
\vspace{-0.5cm}
    \end{center}
  \caption{The CREDO research road map, the ``scenarios'' branch. Here the SHDM decay is used as an example 
  CRE scenario, and 
  for the sake of clarity the products of the decay are represented by just one photon.}
  \label{fig:science-sequence}
\end{wrapfigure}

The key concept of public engagement in CREDO is tightly connected with the big science questions and challenges
that might be undertaken with help of a global cosmic-ray infrastructure. 
What is dark matter? What is the structure of space-time? Is there any observable New Physics 
in cosmic rays correlated on large scale? These
are examples of questions only from the field of astroparticle physics, potentially with much wider impact, e.g. on 
the foundations of science, cosmology, or particle physics.
A sketch of a research road map and chain that is being 
implemented in CREDO is presented in Fig.~\ref{fig:science-sequence}, using a super heavy dark matter (SHDM) 
scenario as an example of verifiable model (for more scientific details see e.g. Ref. ~\cite{ph-credo-cern} and references
therein).
The road map begins with a scenario that predicts CRE: a large size cosmic-ray phenomenon. 
In our example here we point to SHDM, 
one of dark matter scenarios, 
so far not supported but also not excluded by available measurements and observations. 
A global effort to identify
parts of CRE possibly generated in consequence of an SHDM particle decay could help to test this particular scenario, 
which in case of a positive result would be a major scientific breakthrough. 

The super heavy dark matter scenario is just one science case example. In general a CRE capable of generating 
a multi-air-shower event detectable with the available technologies might be born in many ways. 
The most promising models in terms of 
multiplicity of particles or photons in CRE arriving at the Earth are scenarios that involve 
interactions of electromagnetic particles, mostly photons, with 
background radiations or matter encountered during their propagation through the space. The photon-originated 
CRE scenarios might be giving very characteristic multi-air-shower footprints 
composed even of thousands or million air showers (see \cite{niraj-sun-pre}
for an example) spread over the whole globe. Such extremely distributed footprints composed of air showers spanning a wide 
energy range might be out of reach 
for the currently  
operating observatories dedicated only to a specific and typically narrow range of the cosmic energy spectrum at a 
fixed location. The potential spread of the CRE footprint and its energetic range 
determine the observation strategies - they should be based on a joint, global effort and data inter-change of the 
observatories and smaller detectors sensitive to {\it any} cosmic-ray signal. This obviously goes in line with the
current multi-messenger trends of the astroparticle physics, and the cosmic-ray ensembles should be considered as 
a candidate for a new information channel. 

Since we speak about a new, yet not explored channel, the considerations and research concerning
cosmic-ray ensembles should be carried out with a wide perspective, extended over
lower primary energy regimes, a variety of possible interaction types (e.g. including not-so-often
discussed photon splitting
or photon decay - see e.g. Ref.~\cite{photon-decay} for a review) 
and environments (other stars, galactic nuclei, various background radiation,
interstellar matter), even at large astrophysical distances (e.g. even up to $~$1~Gpc in case of the scenario discussed
in \cite{niraj-sun-pre}). 

Successful observations of cosmic-ray ensembles, be it composed of thousands 
or just few photons (or other particles), would offer a unique chance for a new insight into physics in energy regimes 
unavailable in accelerators.

Highlighting 
such a positive and exciting scientific potential proves to be a fair and efficient way of 
engaging the non-expert public. 
This way
the traditional outreach activity gains a new dimension - {\it the scientific purpose} of the public 
participation on a massive scale, with as many private and small devices as possible. Already 
the pure definition of a scientific subject, CRE, determines the necessity of a large scale social engagement. 
It is only a matter of further and deeper investigations to quantify {\it how much} CREDO needs non-experts and how big
can be the difference they can make in which areas of activities within the project. 


Apart from data acquisition, CREDO asks for massive participation of both expert ad non-expert science enthusiasts
to work hand-in-hand towards a better understanding the data acquired by the smartphone cloud within necessarily 
advanced and complex pattern recognition strategies.
As proven in similar projects dealing with massive scale pattern recognition tasks 
(see e.g. \cite{planet-hunters}), a human eye might serve as an indispensable tool to reinforce machine learning. This 
is particularly obvious when the research is oriented on identifying unexpected anomalies in patterns and in the 
signal strength - where no a priori knowledge and training can be applied, and where unsupervised machine 
learning might fail in  getting tuned to the signal features that are ``too strange'' or on borders of categories. 

\section{CREDO status and summary: young, self-standing, inviting openly}
\label{sec:status}

\begin{wrapfigure}{r}{0.5\textwidth}
  \begin{center}
    \vspace{-0.8cm}
    \includegraphics[width=0.48\textwidth]{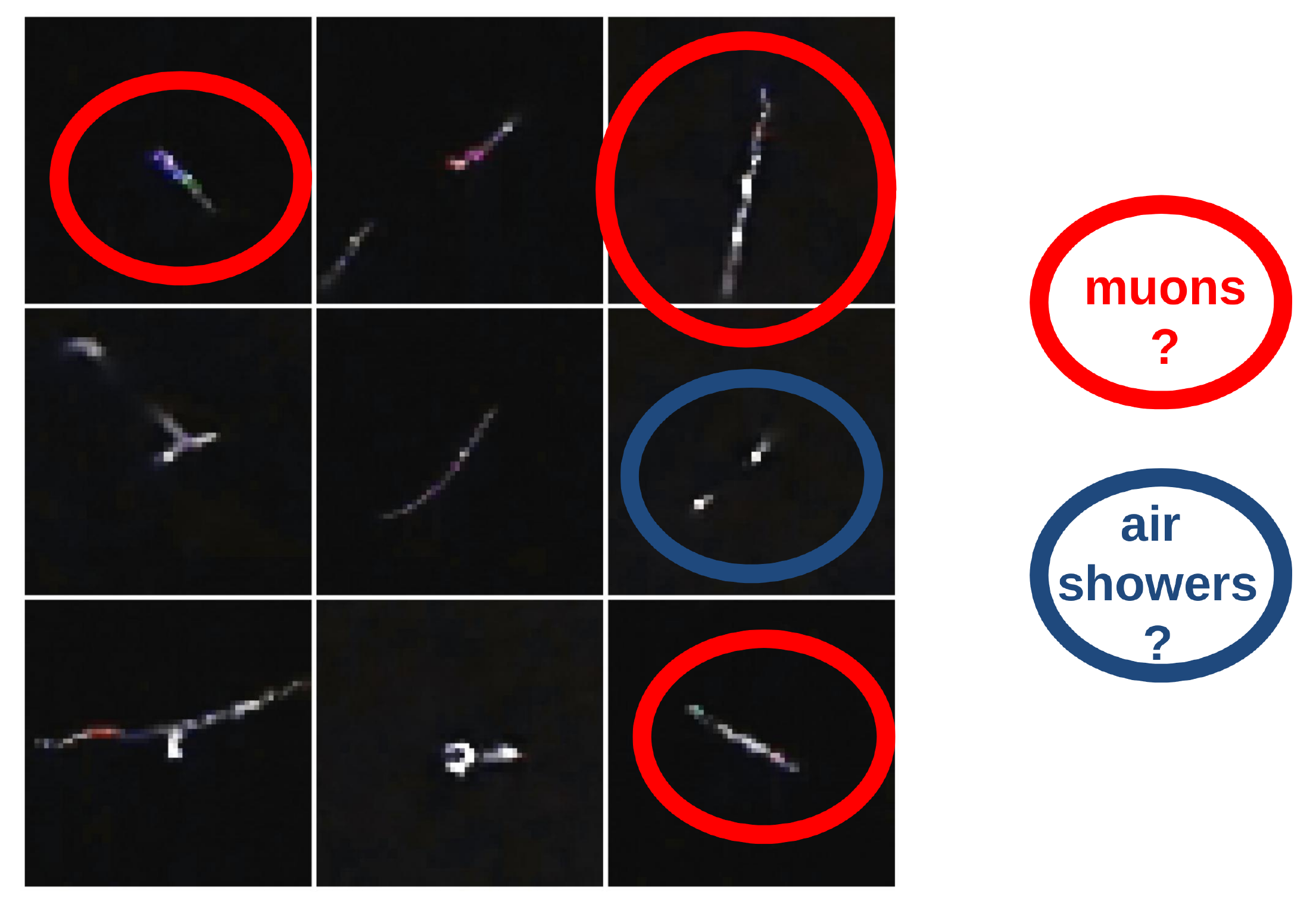}
    \vspace{-0.5cm}
  \end{center}
  \caption{The variety of particle tracks candidates as registered by a smartphone using the CREDO Detector 
app \cite{credo-detector}.}
\vspace{-0.3cm}
  \label{fig:tracks}
\end{wrapfigure}

CREDO is already a self-standing open observatory and international collaboration of 23 institutions from 11 countries 
(\cite{credo-science}.
We have already enabled mass participation data acquisition with smartphones (the CREDO Detector app 
\cite{credo-detector}): > 10k devices,
> 1000 years detection time, > 3M particle track candidates - see Fig. ~\ref{fig:tracks} for examples) and prototyped
software to read data taken by a pocket size detector cloud - adopting Cosmic Watch \cite{cosmic-watch}, 
one of the popular, open-hardware cosmic-ray detection solutions. Thanks to our volunteer participants
we can also offer codes that enable cosmic-ray detection with 
PC/laptop cameras (Windows) and Pi Cameras using Raspberry Pi (Linux). The CREDO data is being stored, 
processed and made available continuously in the CREDO central computing system operated by ACC Cyfronet AGH-UST, 
in Krakow, Poland. The data center 
is open on two ends: every experiment or private detector can submit own data, if it only matches the 
API protocols. Also the access to data is open to the registered users with scientific or educational motivations.
All the CREDO codes are open under the MIT license - so that the community of participants can also get engaged in software 
development - for science and for their education. The key scientific achievement to date
is the launch and first data analysis in Quantum Gravity Previewer - the first mass participation on-line experiment
on the CREDO infrastructure dedicated to search for anomalous clustering cosmic rays in time, potentially a manifestation 
of the space-time structure (\cite{qgp-eurek}). Regarding the already visible societal benefits 
we list e.g. the Particle Hunters program (\cite{particle-hunters} - a component of the CREDO competition systems
that facilitates team competition in particle hunting with smartphones. To date, more than 1200 pupils from 
more than 60 schools in Poland participate in the program with their teachers, 
not only offering their smartphones as the detectors, but also using the educational materials provided by CREDO
(a lesson plan and the accompanying materials, available also in English). 
The popularity and universality of Particle Hunters give prospects for expansion not only on a country level in 
Poland, but also internationally, beginning from the CREDO member countries. The other components of the 
competition system that are already under preparation will further widen the spectrum of scientific activities, 
educational opportunities and simple, positive competitions coupled to international cooperation and contacts available
for the engaged youth. CREDO as a whole will become an attractive, long-term option for young people
hopefully even for a life-long journey with science.

Finally, CREDO is not only an astrophysics project. Once a global, multi-instrument monitoring of cosmic radiation 
is in operation, a number of inter- and trans-disciplinary opportunities will arise, including possible contributions
to biophysics
(e.g. impact of irradiation in central parts of high-energy extensive air showers on human health, following the
trends in low radiation dose research as described e.g. in Ref~\cite{credo-bio}), and 
geophysics (relationships between earthquakes and anomalous changes in the low energy cosmic ray flux reaching the surface 
of our planet - to test the hypotheses discussed e.g. in Ref~\cite{credo-geo}).
The latter would mean a contribution to yet another multi-messenger field, apart from the astrophysical one - 
the {\it inward} multi-messenger physics, looking into the Earth. 
The CREDO door is now open, feel invited, everybody is welcome.\\

\end{document}